\documentclass[prb,preprint,superscriptaddress]{revtex4-1} 

\usepackage{amsmath}  
\usepackage{amsfonts} 
\usepackage{graphicx} 
\usepackage{xcolor}
\usepackage{url}
\usepackage{academicons}
\usepackage{orcidlink}
\newcommand{\orcid}[1]{\href{https://orcid.org/#1}{\textcolor[HTML]{A6CE39}{\aiOrcid}}}

\begin{document}

\title{ 
Design and characterization of a simple polarization grating-based polarimeter}



\author{Massimo Santarsiero}
\email{massimo.santarsiero@uniroma3.it}
\affiliation{DIIEM, Università Roma Tre, via V. Volterra 62, Rome 00146, Italy.}

\author{J. C. G. de Sande}
\email{juancarlos.gonzalez@upm.es}
\affiliation{Universidad Politécnica de Madrid, ETSIS de Telecomunicación, Campus Sur, 28031 Madrid, Spain.}
\author{Gemma Piquero}
\affiliation{Departamento de Óptica, Fac. CC. Físicas, U.C.M., Ciudad Universitaria s/n, 28040 Madrid, Spain.}
\email{piquero@ucm.es}




\date{\today}

\begin{abstract} 

In undergraduate optics courses, diffraction gratings are studied extensively, generally within the scalar approximation. When the vector nature of light is taken into account, so-called polarization diffraction gratings have been proposed, which are a cutting-edge research topic due to their numerous applications. This paper proposes a simple experiment to introduce students to polarization diffraction gratings and, at the same time, use this device to apply many of the concepts learned about polarimetry. Although current research uses spatial light modulators and metasurfaces, we use a cheap commercial polarization grating.
In addition to show how a polarization grating can be characterized, its use as a cheap and easy-to-use Stokes polarimeter is described and demonstrated experimentally. 
In performing the experiment, issues typical of inverting linear systems will arise, and this will also provide the opportunity to address the problem of finding well-conditioned systems of equations.

\end{abstract}

\maketitle 


\section{Introduction}

During the first years of Physics and Engineering degrees, one of the most relevant topics in Optics courses for undergraduate students is diffraction theory and, in particular, the properties and applications of diffraction gratings (DGs) ~\cite{Jenkins01, B&W99,  Hecht02, Pedrotti13}. These devices are used in numerous applications, such as spectrometry, ellipsometry, etc. Due to their importance, many didactical papers related to DGs have also been published~\cite{Martínez:EJP07, Fortin:AJP08, Suhr:EJP11,  Henn:JAP24, Kanchanapusakit:AJP25}.
The study of DGs is usually restricted to the scalar treatment and concerns the properties of the diffracted orders and their dependence on the wavelength of the incident beam and some characteristics of the grating, such as its transmission function and, mostly, its period.

Another highly relevant topic in the study of Optics is polarization, which emerges when the vector nature of light is taken into account~\cite{B&W99, Goldstein03, Chipman18}. Several experiments can be performed in undergraduate laboratories designed to highlight the polarization characteristics of light. In particular, simple schemes have been proposed and used to determine the polarization state of a light beam.~\cite{Salik:AJP12,Patterson:AJP15, Petrova:AJP15, Abayaratne:AJP17}. 

In contrast, experiments that simultaneously involve the diffraction and polarization of light are rather unusual. Instead, as we will show in this work, the two aspects can be profitably combined in a single experiment, highlighting aspects of both. The key element is the so called polarization grating (PG), which diffracts light through a controlled spatial modification of the polarization state, rather than the phase and/or the amplitude of the impinging light, as it happens for standard diffraction gratings.~\cite{Gori:OL99, Tervo:OC01, Davis:OL01, Cincotti:JQE03}. 

Current research on PGs explores their implementation using liquid crystal technologies and dielectric metasurfaces, as well as their integration into practical systems~\cite{Davis:OL01, Emoto:JAP09, Davis:OE16, Rubin:OE18, Liang:ACSAOM24}. It turns out that PGs find applications in very different fields, such as beam steering and splitting~\textcolor{cyan}{\cite{Kim:JLT15, LiZ:OE25}}, compact spectrometers~\textcolor{cyan}{\cite{Qin:OL19}}, polarimetric imaging~\textcolor{cyan}{\cite{Rubin:OE22, Soma:OE24}}, optical communications, augmented and virtual reality displays~\textcolor{cyan}{\cite{Lee:OE17, Yan:OE25}}, and quantum optics experiments~\textcolor{cyan}{\cite{Magallanes:NP18}}. These devices provide a clear example of how polarization, a fundamental property of light, can be harnessed to develop advanced and efficient photonic components. Since Gori introduced this type of grating and proposed it as a tool for performing polarimetry~\cite{Gori:OL99}, numerous scientific articles have been published on the topic, both from a theoretical and experimental point of view, as well as on their design and applications\cite{Tervo:OC01, Piquero:OC01, Davis:OL01,  Badham:JJAP16, Davis:OE16, Rubin:OE18, Liang:ACSAOM24, Li_F:SA25, LiZ:OE25}.

In this paper, we describe an experiment aimed at the realization and characterization of a simple polarimeter, which features a polarization grating as the essential element. As will be shown, in fact, it is sufficient to let a light beam impinge onto a PG and measure the powers of a selected subset of diffraction orders emerging from the grating to completely recover the Stokes parameters of the impinging light (i.e., to perform \emph{Stokes polarimetry}).

In current research PGs are implemented using spatial light modulators~\textcolor{cyan}{\cite{Davis:OE16}} or suitably designed metasurfaces~\textcolor{cyan}{\cite{Rubin:OE18}}. In contrast, here we propose the use of a low-cost commercial PG, the properties of which must first be determined. The latter step, in particular, requires the experimental characterization of the grating, specifically the Mueller matrices associated with its diffraction orders.
This can be done by means of experimental procedures typically employed for characterizing materials or biological samples (\emph{Mueller polarimetry}).

In both Stokes and Mueller polarimetry, one is faced with the problem of  inverting overdetermined linear systems so that, in both phases of the experiment, issues typical of solving ill-conditioned problems arise. Optimal choices have to be made regarding the polarization states to be used for the characterization of the PG, as well as the diffraction orders to be measured to recover the polarization state of the impinging light.

In this way, in a single simple experiment different themes are combined, and the student is faced with several important topics of Physics: diffraction, light polarization (introducing Stokes and Mueller polarimetry), search for well-conditioned systems of equations.  

\section{Preliminaries}

\subsection{Jones formalism and Stokes parameters}
\label{prelminariesA}

A transverse electromagnetic plane wave propagating along the $z$-axis can be expressed, across a given plane $z=$constant, as~\cite{B&W99, Hecht02, Pedrotti13, Goldstein03}
%
\begin{equation}
    \boldsymbol{E}(t)
    = \left( A_{x} \boldsymbol{\hat{x}} + A_{y} \rm {e}^{\rm{i} \varphi} \boldsymbol{\hat{y}}\right) \exp\left[\rm{i}(\omega t+\varphi_0)\right] \, ,
\end{equation}
where $\boldsymbol{\hat{x}}$ and $\boldsymbol{\hat{y}}$ are unit vectors along the $x$ and $y$ axes, respectively, $A_{x}$ and $A_{y}$ are the real amplitudes of each field component, $\omega$ is the angular frequency, $\varphi_0$ is an initial phase and $\varphi$ the phase difference between the two orthogonal components of the field. The values $A_{x}$ and $A_{y}$ together with $\varphi$ define the polarization state of the plane wave and are usually written as a \emph{Jones vector}, that is,
\begin{equation}
\label{Jv0}
    \boldsymbol{E}=\left( \begin{array}{c}
         A_{x}  \\
         A_{y} \, \rm {e}^{\rm{i} \varphi} 
    \end{array} \right) \, .
\end{equation}

The tip of the field evolving in time 
describes an ellipse~\cite{B&W99} across the transverse plane. If $0<\varphi< \pi$ ($-\pi<\varphi< 0$) 
the ellipse is traversed clockwise (counterclockwise) and the polarization is said to be right-handed (left-handed). For the particular case of $\varphi=m \pi$ (with integer $m$), the ellipse degenerates into a segment whose orientation is determined by the ratio $A_{y}/A_{x}$ and the wave is linearly polarized. If $A_{y}=A_{x}$ and the phase difference is $\varphi=\pm \pi/2+m\pi$ the ellipse reduces to a circle and the light is circularly polarized.

 A linear optical system that modifies the polarization state of an incident light beam can be described by means of a complex \emph{Jones matrix},~\cite{Goldstein03}
\begin{equation}
\label{Jonesm0}
\widehat{T} = \left(
    \begin{array}{cc}
        T_{11} &  T_{12}\\
         T_{21} & T_{22}
    \end{array} \right) \, ,
\end{equation} 
and the Jones vector of the output light, $\boldsymbol{E}^{\rm out}$, is related to the Jones vector of the input light, $\boldsymbol{E}^{\rm in}$, as~\cite{Goldstein03}
\begin{equation}
    \boldsymbol{E}^{\rm out}= \widehat{T} \, \boldsymbol{E}^{\rm in} \, .
\end{equation}

The polarization ellipse can also be described by means of the \emph{Stokes parameters}, usually arranged in a four-component \emph{Stokes vector}, which for totally polarized light is defined as~\cite{Goldstein03}
\begin{equation}
    \boldsymbol{S}=\left(
    \begin{array}{c}
         S_0 \\  S_1 \\ S_2 \\ S_3
    \end{array} \right) =  \left( 
    \begin{array}{c} 
         A_{x}^2 + A_{y}^2 \\
         A_{x}^2 - A_{y}^2 \\
         2\, A_{x} A_{y} \, \cos\varphi\\
         2\, A_{x} A_{y} \, \sin\varphi           
    \end{array} \right) \, .
    \label{stokesDefinition}
\end{equation}

\begin{figure}
    \centering
    \includegraphics[width=0.4\linewidth]{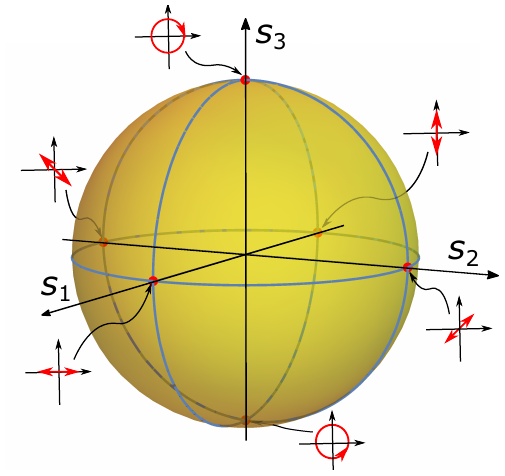}
    \caption{The Poincaré sphere. Are explicitly shown: linearly polarized states along $x$, $y$ and along an angle at $\pm \pi/4$ with respect to $x$, right- and left-handed circularly polarized states.}
    \label{fig:poincare}
\end{figure}

$S_0$ represents the total irradiance, while the other parameters give information about the polarization state of light.  
If we divide the Stokes parameters by $S_0$ we obtain the following \emph{normalized} Stokes vector:
\begin{equation}
    \boldsymbol{s}=
    \left(
    \begin{array}{c}
         S_1/S_0 \\ S_2/S_0 \\ S_3/S_0
    \end{array} \right) = 
     \left(
    \begin{array}{c}
         s_1 \\ s_2 \\ s_3
    \end{array} \right)  \, ,
\end{equation}
which can be visualized in a 3-dimensional space. Since the modulus of $\boldsymbol{s}$ is unitary (for totally polarized light), its tip describes a sphere, the \emph{Poincar\'e Sphere}, each point of which corresponding to a different polarization state. In particular, linear polarization states are represented along the equatorial line, while circular states are at the poles (see Fig.~\ref{fig:poincare}).

The Stokes parameters are real quantities and can be experimentally obtained from measurements of the power content carried by selected polarization components of light. 
Details on these procedures are shown in Appendix A.
We recall here that one possible way to achieve this is by passing the light beam through appropriate sequences of waveplates and/or linear polarizers and measuring the resulting output power. If we denote by $P_x$, $P_y$, $P_{\pm \pi/4}$ the power contents of the linear component along $x$, along $y$ and at $\pm \pi/4$, respectively, and by $P_R$ and $P_L$ the analogous quantities for right- and left-handed circularly polarized components, we obtain 
\begin{equation}
\label{StokesInt}
\left\{
\begin{array}{ll}
S_0 &= \displaystyle\frac{1}{3} \left(P_x + P_y + P_{\pi/4} + P_{-\pi/4}+ P_{R} + P_L \right) = 
\\
& = P_x + P_y = P_{\pi/4} + P_{-\pi/4}=P_R + P_L  \, , \\
S_1 & = P_{x} -P_{y} \, , \\
S_2 & = P_{\pi/4} -P_{-\pi/4}  \, , \\
S_3 & = P_{R} -P_{L} \, . 
\end{array}
\right.
\end{equation}

Equation~\ref{StokesInt} also provides a clearer physical insight into the Stokes parameters. In fact, it turns out that $S_0$ corresponds to the total power of the beam, $S_1$ is the difference between the power of the light content linearly polarized along the $x$ and $y$ axes, $S_2$ is the difference of the power of the light linearly polarized along the bisectrix of the first and fourth quadrants, and $S_3$ is the difference of the powers of the right-handed and left-handed circularly polarized components. 

\subsection{Mueller formalism}
\label{prelminariesB}

When light passes through a linear optical system, the Stokes parameters of incident light, $\boldsymbol{S}^{\rm in}$, and those corresponding to output light, $\boldsymbol{S}^{\rm out}$, are related through a $4 \times 4$ matrix (the \emph{Mueller matrix}) in such a way that~\cite{Goldstein03}
\begin{equation}
\label{Sinout}
    \boldsymbol{S}^{\rm out}= \widehat{M} \; \boldsymbol{S}^{\rm in} \, .
\end{equation}

The Mueller matrix completely characterizes any linear system with respect to its ability to modify the polarization characteristics of light. If a Jones matrix can be associated to a given optical system (and this happens when the system itself is entirely deterministic), the corresponding Mueller matrix can be evaluated as~\cite{Gil16}
\begin{equation}
\widehat{M} = \widehat{\mathcal{L}} \left( \widehat{T} \otimes \widehat{T}^{\ast} \right) \widehat{\mathcal{L}}^{-1} \, ,
\label{Jones2Mueller}
\end{equation}
where the matrix $\widehat{\mathcal{L}}$ is defined as 
\begin{equation}
\widehat{\mathcal{L}}=\left(	
\begin{array}{cccc}
1 & 0 & 0 &1 \\ 1 & 0 & 0 &-1 \\ 0 & 1 & 1 & 0 \\ 0 & {\rm i} & - {\rm i} & 0
\end{array}
\right) \, ,
\end{equation}
and $\otimes$ denotes the Kronecker product\textcolor{magenta}{~\cite{Gil16}}.

The Mueller matrix of a sample can be recovered starting from the relation in Eq.~(\ref{Sinout}), letting light with different polarization states pass through the sample, and measuring the polarization state of emerging light. 
The details of these procedures are given in Appendix B. Here we summarize the main results.

The optimal selection of the input polarization states results when they correspond to the six vertices of an octahedron inscribed in the Poincaré sphere. In particular, we can choose the following states: linearly polarized at 0, $\pi/4$, $-\pi/4$ and $\pi/2$, right- and left-handed circularly polarized. All these states can be easily generated by means of a linear polarizer and a quarter-wave phase plate.

The following two matrices can be built starting from the Stokes parameters of the input ($\boldsymbol{S}^{\rm in,i}$, $i=1,...,6$) and of the output ($\boldsymbol{S}^{\rm out,i}$, $i=1,...,6$)  light:
\begin{eqnarray}
\widehat{S}^{\rm in}=\left(\begin{array}{cccccc}
S_0^{\rm in,1} &
S_0^{\rm in,2} &
S_0^{\rm in,3} &
S_0^{\rm in,4} &
S_0^{\rm in,5} &
S_0^{\rm in,6} \\
S_1^{\rm in,1} &
S_1^{\rm in,2} &
S_1^{\rm in,3} &
S_1^{\rm in,4} &
S_1^{\rm in,5} &
S_1^{\rm in,6} \\ 
S_2^{\rm in,1} & 
S_2^{\rm in,2} &    
S_2^{\rm in,3} &
S_2^{\rm in,4} &
S_2^{\rm in,5} &
S_2^{\rm in,6} \\ 
S_3^{\rm in,1} & 
S_3^{\rm in,2} &
S_3^{\rm in,3} &
S_3^{\rm in,4} &
S_3^{\rm in,5} &
S_3^{\rm in,6} \\
\end{array} \right)  \, ,
\label{MSin}
\end{eqnarray}
and
\begin{eqnarray}
\widehat{S}^{\rm out}=\left(\begin{array}{cccccc}
S_0^{\rm out,1} &
S_0^{\rm out,2} &
S_0^{\rm out,3} &
S_0^{\rm out,4} &
S_0^{\rm out,5} &
S_0^{\rm out,6} \\
S_1^{\rm out,1} &
S_1^{\rm out,2} &
S_1^{\rm out,3} &
S_1^{\rm out,4} &
S_1^{\rm out,5} &
S_1^{\rm out,6} \\ 
S_2^{\rm out,1} & 
S_2^{\rm out,2} &    
S_2^{\rm out,3} &
S_2^{\rm out,4} &
S_2^{\rm out,5} &
S_2^{\rm out,6} \\ 
S_3^{\rm out,1} & 
S_3^{\rm out,2} &
S_3^{\rm out,3} &
S_3^{\rm out,4} &
S_3^{\rm out,5} &
S_3^{\rm out,6} 
\end{array} \right) \, . 
\label{MSout}
\end{eqnarray}

Therefore, an overdetermined system of linear equations can be written as
\begin{eqnarray}
\widehat{S}^{\rm out}&=& \widehat{M} \; \widehat{S}^{\rm in}\, ,
\label{Expi}
\end{eqnarray}
which can be inverted to recover the matrix $\widehat{M}$. This is done (see Appendix B) through the introduction of the pseudo-inverse matrix
\begin{eqnarray}
	\left( \widehat{S}^{\rm in}\right)^{+}&=&\left( \widehat{S}^{\rm in}\right)^{T}\left[ \widehat{S}^{\rm in} \left( \widehat{S}^{\rm in}\right)^T\right]^{-1} \, ,
	\label{M_P_pinv}
\end{eqnarray}
resulting
\begin{equation}
	\widehat{M}=\widehat{S}^{\rm out} 	\left( \widehat{S}^{\rm in}\right)^{+}
	\, .
	\label{Mexp} 
\end{equation}
%

\section{Polarization gratings}
\label{PG}

\begin{figure}
\centering\includegraphics[width=10cm]{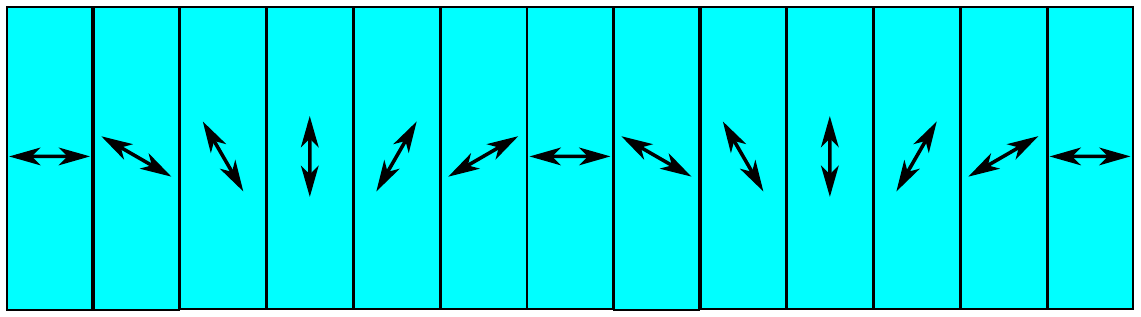}
\caption{Scheme of a simple polarization grating, a transparency in which
the polarization of the incident wave is changed periodically
along a line. Arrows indicate the transmission axis direction of a polarizer.}
\label{Fig1}
\end{figure}

Just as a diffraction grating is characterized by a periodic transmission function, so a polarization grating is associated with a periodic Jones matrix. One of the simplest examples of a polarization grating was proposed by Gori~\cite{Gori:OL99}. It behaves as a linear polarizer, whose transmission direction varies continuously along the $x$ direction. A discretized version of such a grating is sketched in Fig. \ref{Fig1}. 

It can be represented by a spatially varying Jones matrix of the form~\cite{Gori:OL99}
\begin{equation}
\label{goriPG}
\widehat{T}(x)= \left (
    \begin{array}{cc} 
        \cos^2( \pi x/L) 
        &    \cos(\pi x/L ) \,\sin(\pi x/L) \cr
        \cos(\pi x/L) \,\sin(\pi x/L)  &  \sin^2(\pi x/L)
    \end{array} \right),
\end{equation}
where $L$ is the spatial period of the grating. 

Using Euler formulas, the above matrix can be written as 
%
\begin{equation}
\label{goriPGbis}
\begin{array}{rl}
\widehat{T}(x)= 
&
\displaystyle\frac{1}{2} 
    \left (
    \begin{array}{cc} 
        1 & 0  \\
        0 & 1
    \end{array} 
    \right)
    +
\displaystyle\frac{1}{4}
    \left (
    \begin{array}{cc} 
        1 & -{\rm i}  \\
        -{\rm i}  & -1
    \end{array} 
    \right)
    \exp\left({{\rm i}  \, 2 \pi x/L}\right)
    \\
    \\
    &
    + 
\displaystyle\frac{1}{4}
    \left (
    \begin{array}{cc} 
        1 & {\rm i}   \\
        {\rm i}  & -1
    \end{array} 
    \right)
    \exp\left({-{\rm i}  \, 2 \pi x/L}\right)
    \end{array}
\end{equation}
showing that Gori's polarization grating produces only three diffracted orders: the central one (zero order) preserves the polarization state of the incident beam, while the $\pm 1$ orders always produce left- and right-handed circularly polarized light~\cite{Gori:OL99}. 
It has also been shown that such a grating can be used to obtain the Stokes parameters of incident light by measuring the powers of the three diffraction orders after passing through a linear polarizer set with its transmission axis at two different orientations~\cite{Gori:OL99}.

For more general polarization gratings the Jones matrix can be expressed as
\begin{equation}
\label{Jonesx}
\widehat{T} (x) = \left(
    \begin{array}{cc}
        T_{11}(x) &  T_{12}(x)\\
         T_{21}(x) & T_{22}(x)
    \end{array} \right) \, ,
\end{equation}
where $T_{ij}(x)$, $i,j=1,2$, are periodic functions with a common period $L$, so they can be expressed as Fourier series of the form
\begin{equation}
\label{JonesFS}
T_{ij}(x) =
  \sum_{n=-\infty}^{\infty}
	t_{ij}^{(n)} \exp({\rm i}\, 2\, \pi\,  n\, x /L) \, ,
\end{equation}
with
\begin{equation}
\label{JonesFSa}
t_{ij}^{(n)} =\frac{1}{L}
  \int_{0}^{L}
	T_{ij}(x) \exp(-{\rm i}\, 2\, \pi\, n\, x /L) \; {\rm d}x \, .
\end{equation}

Therefore, the Jones matrix of the polarization grating can be written as 
\begin{equation}
\widehat{T}(x) = \sum_{n=-\infty}^{\infty}
	\widehat{T}_{n} \exp({\rm i} \, 2\pi n x /L) \, ,
\label{JonesmFS}
\end{equation}
where
\begin{equation}
\label{Jones_n}
\widehat{T}_n = \left(
    \begin{array}{cc}
        t_{11}^{(n)} &  t_{12}^{(n)}\\
         t_{21}^{(n)} & t_{22}^{(n)}
    \end{array} \right) \, 
\end{equation}
is the Jones matrix corresponding to the $n$-th diffraction order.

In the most general case, the number of diffraction orders, corresponding to the number of terms of the sum in Eq.~(\ref{JonesFS}), is virtually infinite and includes both homogeneous and evanescent waves. However, for our intended use of the grating in polarimetry, it will be sufficient to consider a finite, and generally rather limited, number of homogeneous orders.

For each Jones matrix $\widehat{T}_n$ associated to the $n$-th diffraction order, a corresponding Mueller matrix $\widehat{M}_n$ can be obtained using the transformation given in Eq.~(\ref{Jones2Mueller}), thus obtaining
\begin{equation}
\widehat{M}_{n} = \widehat{\mathcal{L}} \left( \widehat{T}_n \otimes \widehat{T}^{\ast}_n \right) \widehat{\mathcal{L}}^{-1} 
\, .
\label{Jones_Mueller}
\end{equation}
In this way, the polarization grating can be represented by a set of Mueller matrices $\left\{\widehat{M}_{n}\right\}$, $(n=0, \pm 1, \pm 2, \hdots )$, each of which gives account of the property of one diffraction order of modifying the polarization of incident light.

Actually, a complete characterization of the Mueller matrix of the grating would require knowing all the matrices of the form 
\begin{equation}
\widehat{M}_{nm} = \widehat{\mathcal{L}} \left( \widehat{T}_n \otimes \widehat{T}^{\ast}_m \right) \widehat{\mathcal{L}}^{-1} 
\, ,
\label{Jones_Mueller_bis}
\end{equation}
as can be seen by inserting the sum in Eq.~(\ref{JonesmFS}) into Eq.~(\ref{Jones2Mueller}). Nevertheless, since we suppose that the diffraction orders do not overlap when they are detected, the terms with $n\ne m$ do not give any contribution to their polarization .

\section{Experimental characterization of a polarization grating}
\label{PGcharacterization}

In the first part of the experiment we characterize the polarization grating by measuring the Mueller matrices ($\widehat{M}_{n}$) pertaining to its diffraction orders. In view of its application in polarimetry, it is sufficient to examine only a finite number of orders. 
A commercial polarization grating from Edmund Optics (VIS Coated, 25mm Square, 5° Diffraction Polarization Grating) is tested.

\begin{figure}[htbp]
\centering\includegraphics[width=9cm]{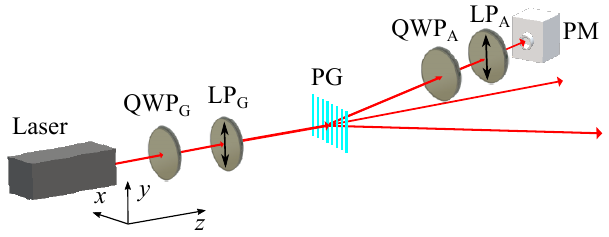}
\caption{Experimental setup. $\rm{QWP_G}$: quarter-wave phase plate;  $\rm{LP_G}$: removable linear polarizer; PG: polarization grating; $\rm{QWP_A}$: removable quarter-wave phase plate;  $\rm{LP_A}$: linear polarizer; PM: power meter.}
\label{setup}
\end{figure}

The technique used is one of the standard ones used for Mueller polarimetry, presented in Appendix~B, which makes use of six input polarization states. The experimental setup is shown in Fig.~\ref{setup}. 

 Light from a He-Ne laser (model 30995, from Research Electro-Optics, $\lambda=632.8$nm, linearly polarized along the vertical direction) is directed towards the polarization grating (PG). Before reaching the grating, the beam passes through a polarization converter, 
which can be set in such a way that the six necessary input polarization states can be generated (namely, linear at 0, $\pi/2$, and $\pm \pi/4$, and right- and left-handed circular). 

Several combinations of linear polarizers and wave plates can be used to realize the polarization converter. In our case, it consists of a quarter-wave phase plate ($\rm{QWP_G}$) and a removable linear polarizer ($\rm{LP_G}$). To produce the two circularly polarized states, the sole quarter-wave plate ($\rm{QWP_G}$) is sufficient, provided that its fast axis is oriented at $\pi/4$ and $-\pi/4$ with respect to the vertical axis. Since a fine angular alignment of the wave plate is quite critical, the rotation by $\pi/2$ to switch between right and left circular polarization can be avoided by simply flipping the plate around the vertical axis. On the other hand, all linear polarization states can be obtained by placing a linear polarizer ($\rm{LP_G}$) just after the wave plate (the latter being oriented at $\pm \pi/4$, indifferently). In such a way, linearly polarized states at 0, $\pi/2$, and $\pm \pi/4$ can be obtained, all with the same power, by suitably rotating the transmission axis of the polarizer.

For every input polarization state, the output Stokes parameters of light propagating along diffraction orders with $-6 \le n \le 6 $ are measured following the approach described in 
Appendix~A by means of an analyzer, consisting of
a quarter-wave plate ($\rm{QWP_A}$), a linear polarizer ($\rm{LP_A}$) and a power meter (PM), as shown in Fig.~\ref{setup}. The distance between the PG and the detector (about 25 cm) is chosen so that the diffracted beams are clearly separated from one another.

By constructing the matrices of Eqs.~(\ref{MSin}) and (\ref{MSout}) from the values of the measured Stokes parameters, and using Eq.~(\ref{Mexp}), the Mueller matrices $\widehat{M}_n$ were determined for diffracted orders with $-6 \le n \le 6$ (see Tables~\ref{Mn03},~\ref{Mn34}, and ~\ref{Mn46} in Appendix C). 

\begin{figure}[h]
\centering
\includegraphics[width=10cm]{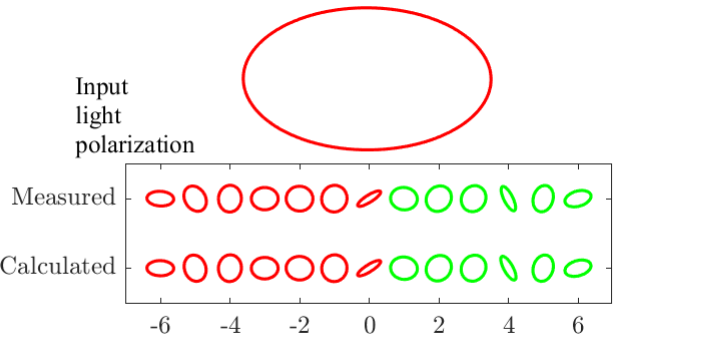}
\caption{Measured polarization ellipse of the input light and those corresponding to different diffraction orders. The calculated polarization ellipses are also shown for comparison. Red (green) ellipses denote right-handed (left-handed) polarization.}
\label{E30}
\end{figure}

At this stage, it is possible to check the correctness of the values obtained for $\widehat{M}_n$ by letting a different polarization state impinge onto the grating and measuring the polarizations of some of the diffraction orders. The check has been performed using an incident elliptical polarization (upper part of Fig.~\ref{E30}), measuring the Stokes parameters of orders with $-6 \le n \le 6$, and comparing the resulting polarization states with those obtained numerically from the matrices $\widehat{M}_n$.
The results, reported in the lower part of Fig.~\ref{E30}, show excellent agreement between the measured polarization states and the calculated ones.
%

\section{Polarization grating as a Stokes polarimeter}

Every instrument aimed at characterizing the polarization of light is generically called a Stokes polarimeter. 
The most common technique used for this task is the one described after Eq.~(\ref{StokesInt}), when the definition of the Stokes parameters was given, and in Appendix~A. Following that approach, a certain number of values of the light power have to be measured, obtained by letting light pass through a given sequence of anisotropic optical elements. In traditional Stokes polarimeters, such measurements are performed in sequence.

A single PG can be used as a Stokes polarimeter. The basic idea is that the powers carried by different diffraction orders can provide the required information to recover the polarization state of the light passing through the grating. 
Thus, Stokes polarimetry is achieved solely via the PG (without further anisotropic optical elements), enabling all necessary measurements to be taken in a single shot.

Current research in this field concerns the design and fabrication of diffracting structures (generally within the framework of metasurfaces) having prescribed features such as to give rise to diffraction orders with powers directly relatable to the Stokes parameters of the incident light. Here, we propose the use of a cheap commercial diffraction polarization grating (by Edmund Scientific), which is not sold for polarimetric purposes but can act a Stokes polarimeter, provided that it has been properly characterized. The principles of the method are presented here.

Let us denote by $m_{ij}^{(n)}$, $(i,j=0,1,2,3)$, the typical element of the Mueller matrix $\widehat{M}_{n}$. If light with unknown polarization state (identified by the Stokes vector ${\boldsymbol S}^{\rm in}$) is sent onto the grating, the power carried by the $n$-th diffraction order, $P_n^{\rm out}$, is given by the first element of the corresponding Stokes vector, ${\boldsymbol S}^{{\rm out},n}$, that is
\begin{equation}
    P_n^{\rm out}= 
    S_0^{{\rm out}, n} =
    m_{00}^{(n)}S_0^{\rm in} +
        m_{01}^{(n)}S_1^{\rm in} +
        m_{02}^{(n)}S_2^{\rm in} +
        m_{03}^{(n)}S_3^{\rm in} \, .
\end{equation}

We can arrange the measured powers of the diffracted orders (with $-N \le n \le N$) in the column vector $\boldsymbol{P}^{\rm out}$, so that the following relation can be written:
\begin{equation}
  \boldsymbol{P}^{\rm out}=  
  \widehat{M}_{M}  
  \boldsymbol{S}^{\rm in} \, ,
  \label{PMS}
\end{equation}
where a $4 \times (2N+1)$ measurement matrix $\widehat{M}_{M}$ has been built by stacking the first row of the $\widehat{M}_n$ Mueller matrices, i.e., 
\begin{eqnarray}
\label{Pown}
\widehat{M}_{M}=
      \left(
    \begin{array}{cccc}
        m_{00}^{(-N)}  &  m_{01}^{(-N)} &  m_{02}^{(-N)} & m_{03}^{(-N)}  \\
        \vdots & \vdots & \vdots & \vdots\\
        m_{00}^{(0)}  &  m_{01}^{(0)} &  m_{02}^{(0)} & m_{03}^{(0)} \\
        \vdots & \vdots & \vdots & \vdots\\
        m_{00}^{(N)}  &  m_{01}^{(N)} &  m_{02}^{(N)} & m_{03}^{(N)} 
    \end{array}
    \right)  
\; ,
\end{eqnarray}
and the problem reduces to inverting Eq.~(\ref{PMS}) to recover the unknown vector $\boldsymbol{S}^{\rm in}$.

If the number of measured powers exceeds four, the number of equations is greater than the number of unknowns and the linear system turns out to be overdetermined. The problem to be solved is the same as that presented in Appendix A, where the standard technique to recover the Stokes parameters of light was presented.
Even in this case, necessary condition for the problem to be solved is that the rank of $\widehat{M}_{M}$ is four, that is, the columns of $\widehat{M}_{M}$ are linearly independent. This condition, in principle, is not guaranteed because it depends on the values of the elements of the Mueller matrices $\widehat{M}_{n}$, and must be checked starting from the measured matrix elements.

When the condition is fulfilled, the method of the Moore-Penrose pseudoinverse matrix (see Appendix A) can be used.~\cite{Penrose_1955}
Therefore, defining 
\begin{eqnarray}
	\widehat{M}_{M}^{+}
    &
    =
    &
    \widehat{M}_{M}^{T}\left( \widehat{M}_{M} 
    \widehat{M}_{M}^T\right)^{-1} \, ,
	\label{M_P_pinvM}
\end{eqnarray}
the least-squares solution can be expressed as
\begin{equation}
	\boldsymbol{S}^{\rm in}=
    \widehat{M}_{M}^{+} 
    \;
    \boldsymbol{P}^{\rm out}
	\, .
	\label{Mexpb} 
\end{equation}

We also recall what is reported in Appendix~A, that is, to prevent the amplification of errors, the matrix to be inverted, besides being non-singular, is also requested to be well-conditioned. In other terms, it must have a sufficiently low condition number~\cite{cheney2007numerical}.

Unfortunately, by using all detected diffraction orders ($-6\le n \le 6$), the matrix $\widehat{M}_{M}$ turns out to be ill-conditioned (its condition number being of the order of $2\times 10 ^3$), meaning that its pseudoinverse would yield Stokes parameters recovered with large errors. 

To overcome this problem, an appropriate set of rows (i.e., of diffraction orders) can be chosen in such a way that the condition number of the resulting measurement matrix is minimized and the errors on the recovered Stokes parameters are minimized as well. 
Among all possible choices, it is found that by selecting diffracted beams with orders $n=\pm 1$, $n= \pm 3 $, and $n=\pm 4 $, the condition number of the matrix $\widehat{M}_{M}$ results as low as 5.7. Using this set of output powers, the corresponding optimal measurement matrix turns out to be
\begin{eqnarray}
  \widehat{M}_{M}^{\rm opt} =
  \left(
  \begin{array}{cccc}
      0.650\pm 0.004&  0.149\pm 0.007&  -0.171\pm 0.010&  0.242\pm 0.010 \\
      3.75\pm 0.02  &  0.07\pm 0.05  &   0.68\pm 0.04  & -3.63\pm 0.03  \\
      3.95\pm 0.02  &  0.79\pm 0.05  &  -1.32\pm 0.06  & -3.41\pm 0.03  \\
      3.94\pm 0.03  & -0.44\pm 0.06  &   1.72\pm 0.04  &  3.14\pm 0.08  \\
      3.34\pm 0.02  &  0.03\pm 0.04  &  -0.28\pm 0.05  &  2.99\pm 0.07  \\
      7.67\pm 0.05  & -6.65\pm 0.15  &  -0.33\pm 0.10  & -0.20\pm 0.09
  \end{array}
  \right) \times 10^{-4} \, ,
\end{eqnarray}
and can be used to recover the polarization state of an input light beam.

Two different polarization states (pol.~1 and pol.~2) have been generated to test the method. Their Stokes parameters have been measured using the standard six-measure technique presented in Appendix~A (St), and from the powers of the orders diffracted by the PG with $n=\pm 1, \pm 3, \pm 4$ (PG). The results obtained are the following:

\begin{eqnarray} 
\begin{array}{lcl}
\boldsymbol{S}^{\rm 1, st} 
& = & 
3.10 \cdot \left(1.000\pm 0.010, \,0.505\pm 0.006, \, -0.005\pm 0.002, \,  0.865 \pm 0.009 \right) ^T \; ,\\
\boldsymbol{S}^{\rm 1, PG} 
& = &
3.13 \cdot \left(1.000\pm 0.011, \,0.53 \pm 0.03,\, 0.00\pm 0.04,\,  0.892\pm 0.014 \right) ^T \; , 
\end{array}
\label{results1}
\end{eqnarray}
and
\begin{eqnarray} 
\label{results2}
\begin{array}{lcl}
\boldsymbol{S}^{\rm 2, st}
& = & 
   1.90 \cdot \left(1.000\pm 0.010,\, -0.510\pm 0.005,\,  0.004\pm 0.002,\, -0.860\pm 0.008 \right) ^T \; ,
   \\
\boldsymbol{S}^{\rm 2, PG}
& = & 
1.92 \cdot \left(1.000\pm 0.014, \,-0.41 \pm 0.04,\, -0.04 \pm 0.04,\, -0.87\pm 0.02 \right) ^T
\; ,
\end{array}
\end{eqnarray}
respectively.

\begin{figure}[htbp]
\centering\includegraphics[width=8cm]{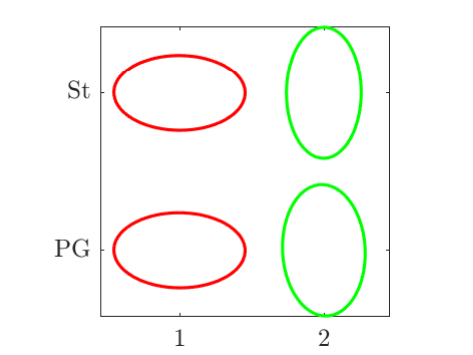}
\caption{Measured polarization (St) ellipse of the input light and calculated polarization (PG) ellipse from the measured powers for diffracted orders $n=\pm 1$, $n= \pm 3 $ and $n=\pm 4 $. Red (green) ellipses denote right-handed (left-handed) polarization.}
\label{MCE30}
\end{figure}

Fig.~\ref{MCE30} shows the polarization ellipses corresponding to the above Stokes vectors.
As can be seen from a comparison of the Stokes parameters and simply by looking at Fig.~\ref{MCE30}, a very good agreement is achieved between them. 

\section{Conclusions}

In this work, polarization diffraction gratings have been introduced to students of Physics and Engineering courses through an experiment that can be carried out in an undergraduate optics laboratory, using a cheap commercial PG. The experiment is aimed at the characterization of the PG through the Mueller matrices associated with its diffraction orders. But students can go further and use it as a simple Stokes polarimeter. In fact, polarization of an input beam can be obtained directly through power measurements of selected diffracted orders.
In this way, students are confronted with two current research topics with numerous applications: diffraction and polarization. Moreover, the experiment will also show common problems encountered when linear systems have to be inverted and lets students explore how well-conditioned equations can be found.


\begin{acknowledgments}

This work has been supported by Spanish Ministerio de Econom\'ia y Competitividad under project PID2023-148021NB-I00.

\end{acknowledgments}

\section*{AUTHOR DECLARATIONS}

The authors have no conflicts to disclose.


\section*{Appendix A. Measuring Stokes parameters of light}
\label{StokesPol}

Techniques aimed at determining the polarization of light beams fall within the so-called Stokes polarimetry. Characterization is achieved through the measurement of its Stokes parameters. It is reasonable that, since the latter are four real quantities, four suitably performed measurements could be sufficient to recover them completely. This is actually the case. 

One of the simplest set of measurable quantities useful for recovering the Stokes parameters of a beam consists of the four powers that are detected by letting the beam pass through four optical elements intended to select, in each case, a particular polarization component of the incident beam.~\cite{B&W99, Goldstein03}

The first quantity to be measured is the power, say $P_x$, of the $x-$component of the electric field, which can be selected by letting the beam pass through a linear polarizer whose transmission axis is oriented along $x$. The second and the third are analogous powers, but pertinent to the $y-$axis ($P_y$) and to an axis oriented at $\pi/4$ with respect to the $x-$axis ($P_{\pi/4}$). The last quantity is the contribution to the total power of the right-handed circularly polarized component ($P_{R}$), which can be measured by letting the beam pass through  a quarter-wave plate (with its fast axis oriented along $x$), followed by a linear polarized with its axis at $\pi/4$.

Using the definition of the Stokes parameters given in Eq.~(\ref{stokesDefinition}), the following relation can be derived 
\begin{equation}
\label{Pxy4Rprima}
    \left( \begin{array}{c}
         P_x  \\
         P_{\pi/4} \\
         P_y  \\
         P_R
    \end{array} \right)=
    \displaystyle\frac{1}{2}
    \left( \begin{array}{cccc}
        1 & 1 & 0 & 0  \\
         1 & 0 & 1 & 0  \\
         1 & -1 & 0 & 0  \\
         1 & 0 & 0 & 1  
    \end{array} \right)
        \left( \begin{array}{c}
         S_0  \\
         S_1 \\
         S_2 \\
         S_3
    \end{array} \right) \, .
\end{equation}
which can be inverted to obtain the Stokes vector of the input light 
\begin{equation}
\label{Pxy4Rseconda}
        \left( \begin{array}{c}
         S_0  \\
         S_1 \\
         S_2 \\
         S_3
    \end{array} \right) 
    =
    \left( \begin{array}{cccc}
        1 & 0 & 1 & 0  \\
         1 & 0 & -1 & 0  \\
         -1 & 2 & -1 & 0  \\
         -1 & 0 & -1 & 2  
    \end{array} \right)
    \left( \begin{array}{c}
         P_x  \\
         P_{\pi/4} \\
         P_y  \\
         P_R
    \end{array} \right)
\, ,
\end{equation}
or, more explicitly,
\begin{equation}
\label{Pxy4Rterza}
\left\{
\begin{array}{l}
S_0 = P_x + P_y   \, , \\
S_1 = P_{x} -P_{y} \, , \\
S_2 = 2 P_{\pi/4} -(P_x + P_y)   \, , \\
S_3 = 2 P_{R} -(P_x + P_y) \, .
\end{array}
\right.
\end{equation}
Now, it is important to point out that inverting a matrix, although non-singular, whose elements are affected by experimental errors, (as in our case, due to errors in the orientation of the element axes, non-ideal behavior of the elements, experimental errors in the power measurements), might result in inaccurate results, as a consequence of error propagation. 

To better understand this point, consider the simplest case of a matrix $2 \times 2$. If the two columns of the matrix define two orthogonal vectors of the same length, the inversion reduces to a scaling factor, and the propagated relative errors are equal to the original ones. In fact, we can choose the reference frame in such a way that the two vectors are aligned with the axes and the matrix transformation turns out to be proportional to the identity matrix, which coincides with its inverse up to a scaling factor. However, if the two vectors have the same length (for example, unitary) but are almost parallel (say, along $x$), very large errors will be made when trying to determine the $y$ component of the solution. Something similar occurs if the two vectors are orthogonal, but when one is much longer than the other. 

This behavior cannot be ascribed to a value of the determinant close to zero, as could be expected. In fact, the following two matrices:
\begin{equation}
	\widehat{A}=
    \left(
    \begin{array}{cc}
    1 & 0.9998 \\
    0 & 0.02 
    \end{array}
    \right)
	\, ,
    \hspace{1cm}
    \widehat{B}=
    \left(
    \begin{array}{cc}
    100 & 0 \\
    0 & 1 
    \end{array}
    \right)
    \, ,
	\label{matriciA} 
\end{equation}
which correspond to the two examples mentioned above, have very different determinants (0.02 and 100, respectively) but lead to the same problem.
Instead, to prevent the amplification of errors, the matrix to be inverted $\widehat{M}$, besides being non-singular, is also requested to be \emph{well-conditioned}, that is, it must have a sufficiently low \emph{condition number}~\cite{cheney2007numerical}. The latter is given by
\begin{equation}
	\kappa(\widehat{M})=||\widehat{M}||\cdot ||\widehat{M}^{-1}|| 
	\, ,
	\label{condNA} 
\end{equation}
where $||\cdot||$ represents the norm of the matrix, which can be defined (among other possibilities) as the Euclidean or $\ell_2$-norm, that is, its largest singular value~\cite{cheney2007numerical}.
While for a matrix proportional to the identity matrix the definition in Eq.~(\ref{condNA}) gives $\kappa=1$, for the two matrices in Eq.~(\ref{matriciA}) it gives $\kappa(\widehat{A})\simeq \kappa(\widehat{B})=100$.

\begin{figure}
    \centering
    \includegraphics[width=0.3\linewidth]{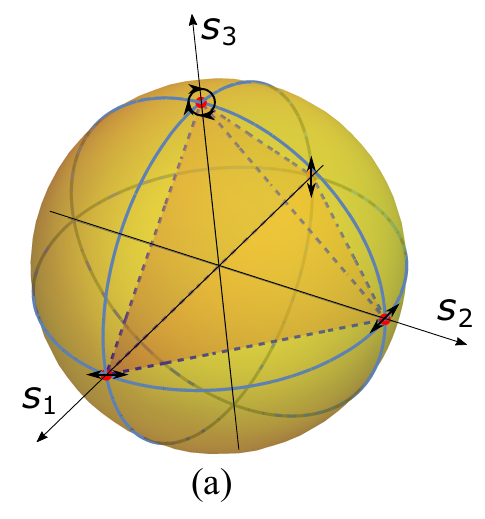}
    \quad
    \includegraphics[width=0.3\linewidth]{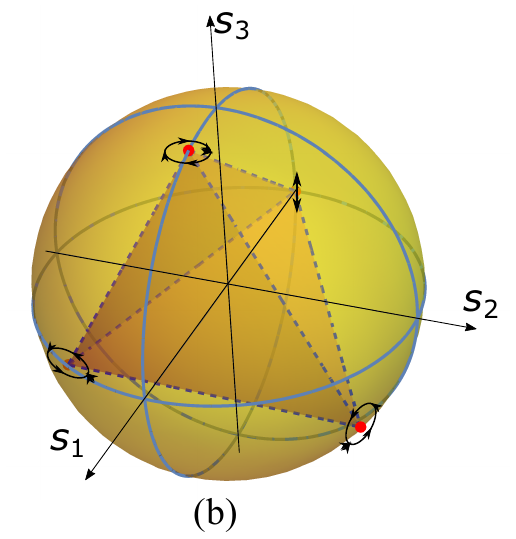}
    \quad
    \includegraphics[width=0.3\linewidth]{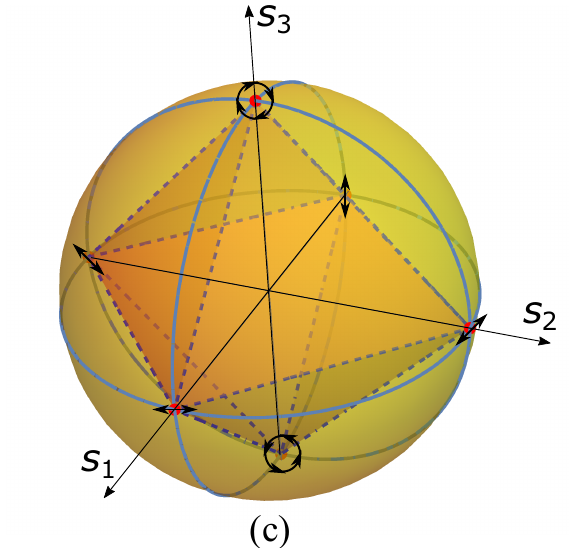}
    \caption{
    (a) Poincaré sphere and an inscribed irregular tetrahedron. (b) Poincaré sphere and an inscribed  regular tetrahedron. (c) Poincaré sphere and an inscribed regular octahedron.}
    \label{Poincare_T_Oct}
\end{figure}

Returning to the Stokes-parameter determination, the matrix in Eq.~(\ref{Pxy4Rprima}) is non-singular but has a relatively high condition number ($\simeq 3.23$). Basically, this stems from a sub-optimal distribution of the sampled polarization states across the Poincaré sphere (see Fig.~\ref{Poincare_T_Oct} (a)). 
A lower condition number would be achieved by choosing the states analyzed more uniformly across the Poincaré sphere, as in Fig.~\ref{Poincare_T_Oct} (b). 
It can be shown, in fact, that an optimum selection of the analyzed components results when the Stokes vectors of the latter correspond to the vertices of a tetrahedron inscribed on the Poincaré sphere. In such a case, the condition number turns out to be $\kappa=\sqrt{3}\simeq 1.73$.

The main drawback of using an optimal distribution of polarization states across the Poincaré sphere is the complexity involved in their synthesis.
Specifically, with the exception of one state (the linear polarization along the vertical axis in Fig.~\ref{Poincare_T_Oct} (b)), the remaining three require specific combinations of precisely oriented waveplates and polarizers for accurate characterization.

An approach to reduce the condition number while maintaining the use of easily implementable polarization analyzers consists of increasing the total number of power measurements. A common choice 
of the states analyzed is the one shown in Fig.~\ref{Poincare_T_Oct} (c). 
In this case, six measurements are used: $P_x$, $P_y$, and $P_{\pm \pi/4}$, which are the powers of linearly polarized components at $0$, $\pi/2$, and $\pm \pi/4$, respectively, and can be measured by letting the beam pass through a linear polarizer oriented along suitable directions. Furthermore, the combination of a quarter-wave plate, with its fast axis along the $x$ direction, followed by a linear polarizer with transmission axis at $\pm \pi/4$ can be used to measure the power of the right-handed ($P_R$) and left-handed ($P_L$) circularly polarized components~\cite{B&W99, Goldstein03}.

The six measured powers can be related to the Stokes parameters through Eq.~(\ref{stokesDefinition}), and the result is
\begin{equation}
\label{Pxy46prim}
    \boldsymbol{P}=
    \widehat A \;
     \boldsymbol{S} \, ,
\end{equation}
with
\begin{equation}
\label{Pxy46prima}
\boldsymbol{P}
=
    \left( \begin{array}{c}
         P_x  \\
         P_{\pi/4} \\
         P_y  \\
         P_{-\pi/4} \\
         P_R\\
         P_L
    \end{array} \right)
\hspace{1 cm}
{\rm and}
\hspace{1 cm}
\widehat A=
    \displaystyle\frac{1}{2}
    \left( \begin{array}{cccc}
        1 & 1 & 0 & 0  \\
         1 & 0 & 1 & 0  \\
         1 & -1 & 0 & 0  \\
         1 & 0 & -1 & 0  \\
         1 & 0 & 0 & 1  \\
         1 & 0 & 0 & -1  
    \end{array} \right)
    \; .
\end{equation}

Now the number of measured powers exceeds four, so that the number of equations is greater than the number of unknowns, and the linear system turns out to be overdetermined. 
A necessary condition for inverting $\widehat{A}$ is that its rank is four, that is, its columns must be linearly independent. This comes from the fact that a submatrix obtained by selecting only four rows of $\widehat A$ must be sufficient to solve the problem and its determinant cannot be zero.
It should be said that such a condition, in principle, is not guaranteed for an arbitrary choice of the analyzed states, but in the present case it is fulfilled.
In the presence of noisy data, the system cannot be solved exactly, and selecting different sets of rows would give rise to different values of the Stokes vector. A simple technique, which takes all the rows into account and provides a least-squares solution to the problem, makes use of the so called \emph{Moore-Penrose pseudoinverse} matrix, defined as~\cite{Penrose_1955}
\begin{eqnarray}
	\widehat{A}^{+}
    &
    =
    &
    \widehat{A}^{T}\left( \widehat{A} 
    \;
    \widehat{A}^T\right)^{-1} \, ,
	\label{M_P_pinvA}
\end{eqnarray}
where $T$ denotes transposed. $\widehat{A}^{+}$ reduces to $\widehat{A}^{-1}$ for square matrices but, in general, enables the system's solution to be expressed as
\begin{equation}
        \boldsymbol{S}
             =
    \widehat{A}^{+} 
    \boldsymbol{P}
\; ,
	\label{Mexpc} 
\end{equation}
with
\begin{equation}
\label{Pxy46seconda}
\widehat{A}^{+} 
=
    \left( \begin{array}{cccccc}
        1/3 & 1/3 & 1/3 & 1/3 & 1/3 & 1/3  \\
         1 & 0 & -1 & 0 & 0 & 0 \\
         0 & 1 & 0 & -1 & 0 & 0 \\
         0 & 0 & 0 & 0 & 1 & -1
    \end{array} \right)
    \; .
\end{equation}
or, more explicitly,
\begin{equation}
\label{StokesIntbis}
\left\{
\begin{array}{l}
S_0 = \displaystyle\frac{1}{3} \left(P_x + P_y + P_{\pi/4} + P_{-\pi/4}+ P_{R} + P_L \right) \,  \\
S_1 = P_{x} -P_{y} \,  \\
S_2 = P_{\pi/4} -P_{-\pi/4}   \\
S_3 = P_{R} -P_{L} \, .
\end{array}
\right.
\end{equation}

The condition number can be evaluated from Eq.~(\ref{condNA}), provided that the inverse of the matrix is replaced by the pseudoinverse and, for the matrix $\widehat{A}$ it turns out to be  $\sqrt{3}\simeq 1.73$.

Given an experimental measurement matrix $\widehat{A} \pm \widehat{\sigma}_{A}$, the propagated errors for each element of the inverse (or pseudoinverse) matrix $A_{pq}^{-1}$, can be estimated as~\cite{Lefebvre:NIMPR00}
\begin{equation}
    (\sigma_{A^{-1}})_{pq}=
        \sqrt{\sum_{i,j} 
        \left(A^{-1}_{pi}\right)^2 
        (\sigma_A)^2_{ij} 
        \left(A^{-1}_{jq}\right)^2} \, .
\end{equation}
In a similar way, the propagated errors for each element of the product of two measured matrices $\widehat{A} \pm \widehat{\sigma}_{A}$ and $\widehat{B} \pm \widehat{\sigma}_{B}$, can be estimated as
\begin{equation}
    (\sigma_{AB})_{pq}=\sqrt{\sum_{i}(A_{pi})^2 (\sigma_B)^2_{iq} +
    (\sigma_A)^2_{pi} (B_{iq})^2} \, .
\end{equation}

\section*{Appendix B. Measuring the Mueller matrix of a sample}
\label{MuellerPol}

Techniques aimed at measuring the Mueller matrix of a sample  belong to the field of so-called Mueller polarimetry.
They consist in letting light with different, known polarization states pass through the sample and measuring the polarization state of emerging light. From a mathematical point of view, they require the inversion of the following relation:
\begin{equation}
\label{SinoutApp}
    \boldsymbol{S}^{\rm out}= \widehat{M} \; \boldsymbol{S}^{\rm in} \, ,
\end{equation}
to recover $\widehat{M}$. 
The problem is analogous to the one encountered when the Stokes parameters of a light beam have to be measured (see Appendix A). 

The minimum number of different polarization states necessary to invert Eq.~(\ref{SinoutApp}) is four. In fact, starting from the Stokes parameters of the input ($\boldsymbol{S}^{\rm in,i}$, $i=1,...,4$) and of the output ($\boldsymbol{S}^{\rm out,i}$, $i=1,...,4$) light, the following two polarimetric measurement matrices can be built :
\begin{eqnarray}
\widehat{S}^{\rm in}=\left(\begin{array}{cccc}
S_0^{\rm in,1} &
S_0^{\rm in,2} &
S_0^{\rm in,3} &
S_0^{\rm in,4} 
\\
S_1^{\rm in,1} &
S_1^{\rm in,2} &
S_1^{\rm in,3} &
S_1^{\rm in,4}
 \\ 
S_2^{\rm in,1} & 
S_2^{\rm in,2} &    
S_2^{\rm in,3} &
S_2^{\rm in,4}
\\ 
S_3^{\rm in,1} & 
S_3^{\rm in,2} &
S_3^{\rm in,3} &
S_3^{\rm in,4}
 \\
\end{array} \right)  \, ,
\label{MSinAp}
\end{eqnarray}
and
\begin{eqnarray}
\widehat{S}^{\rm out}=\left(\begin{array}{cccc}
S_0^{\rm out,1} &
S_0^{\rm out,2} &
S_0^{\rm out,3} &
S_0^{\rm out,4} \\
S_1^{\rm out,1} &
S_1^{\rm out,2} &
S_1^{\rm out,3} &
S_1^{\rm out,4} \\ 
S_2^{\rm out,1} & 
S_2^{\rm out,2} &    
S_2^{\rm out,3} &
S_2^{\rm out,4} \\ 
S_3^{\rm out,1} & 
S_3^{\rm out,2} &
S_3^{\rm out,3} &
S_3^{\rm out,4} 
\end{array} \right) \, ,
\label{MSoutAp}
\end{eqnarray}
so that the matrix $\widehat{M}$ turns out to be
\begin{equation}
	\widehat{M}=\widehat{S}^{\rm out} 	\left( \widehat{S}^{\rm in}\right)^{-1}
	\, .
	\label{MexpAp} 
\end{equation}
provided that the matrix $\widehat{S}^{\rm in}$ is non-singular, that is, that the input Stokes vectors are linearly independent.

When noise is present in the measured data, the four input Stokes vectors, in addition of being independent, have to be chosen in such a way that the reconstruction error on the elements of the Mueller matrix is minimized.
In such a case, an optimum selection of the input polarization results when the Stokes vectors of the input beams correspond to the vertices of a tetrahedron inscribed on the Poincar\'e sphere~\cite{Azzam:JOSAA88,Ambirajan:OEn95, Layden:OE12, Suarez:OLEN19}, as the one shown in Fig.~\ref{Poincare_T_Oct} (a). Unfortunately, at least three of the points correspond to elliptical polarization states that are not easy to generate in the laboratory. 

Nevertheless, an overdetermined system of linear equations results if more than four input polarization states are used. Such a choice makes the results more reliable and less sensitive to noise. For the case of six input polarization states, an optimum selection is obtained when the Stokes vectors of the input beams correspond to the vertices of an octahedron inscribed on the Poincar\'e sphere~\cite{Layden:OE12}. This has the further advantage that the input polarization states can be selected as linearly polarized at 0, $\pm \pi/4$ and $\pi/2$, together with right- and left-handed circularly polarized light. These polarization states are the vertices of an octahedron shown in Fig.~\ref{Poincare_T_Oct} (b), and can be easily generated by means of a linear polarizer and a quarter-wave phase plate.

Now the polarimetric measurement matrices have dimension $4 \times 6$ and the right Moore-Penrose pseudo-inverse~\cite{Penrose_1955} of the matrix $\widehat{S}^{\rm in}$, namely,
\begin{eqnarray}
	\left( \widehat{S}^{\rm in}\right)^{+}&=&\left( \widehat{S}^{\rm in}\right)^{T}\left[ \widehat{S}^{\rm in} \left( \widehat{S}^{\rm in}\right)^T\right]^{-1} \, ,
	\label{M_P_pinvAp}
\end{eqnarray}
can be used instead of the inverse to invert Eq.~(\ref{SinoutApp}), yielding
\begin{equation}
	\widehat{M}=\widehat{S}^{\rm out} 	\left( \widehat{S}^{\rm in}\right)^{+}
	\, .
	\label{MexpApMP} 
\end{equation}
%

\section*{Appendix C: Measured Mueller matrices for several diffraction orders produced by the PG}

Tables \ref{Mn03}, \ref{Mn34}, and \ref{Mn46}, below, show the Mueller matrices corresponding to each diffracted order with $-6 \le n \le 6$. 
\begin{table}[!ht]
\caption{Mueller matrices measured for each diffraction order}
  \label{Mn03}
  \centering
  \begin{ruledtabular}
\begin{tabular}{ccc}
Diffracted order & Normalized Mueller matrix $\widehat{M}_k/m_{00}^{(n)}$  & $m_{00}^{(n)} \times 10^4$\\
\hline
\\
$n = 0$ & $\left(\begin{array}{cccc}
     1.000 \pm 0.006 &   0.235 \pm 0.011 &  
            -0.043 \pm 0.013 &  -0.287 \pm 0.010 \\
    0.110 \pm 0.003  &  0.340 \pm 0.005  & 
            -0.867 \pm 0.011  &  0.050 \pm 0.004 \\
    0.165 \pm 0.003 &  0.881 \pm 0.009  &  
            0.325 \pm 0.004 &  0.087 \pm 0.005 \\
   -0.310 \pm 0.003  & -0.172 \pm 0.005  &  
        0.028 \pm 0.006  &  0.956 \pm 0.008
\end{array}\right)$ 
&  $157$  \\  \\ 
$n = 1$ & $\left(\begin{array}{cccc}
     1.000 \pm 0.007 &  -0.112 \pm 0.014 &   0.436 \pm 0.010&   0.80 \pm 0.02 \\
    0.178 \pm 0.013  & -0.069 \pm 0.003 &   0.048\pm 0.002  &  0.157 \pm 0.004\\
   -0.007\pm 0.001  & -0.044 \pm 0.001 &  0.023\pm 0.001  & -0.035 \pm 0.001\\
   -0.964 \pm 0.007  &  0.106 \pm 0.014 &  -0.437 \pm 0.010  & -0.79\pm 0.02
\end{array}\right)$
&  $3.94$  \\ \\ 
$n = -1$ & $\left(\begin{array}{cccc}
    1.000\pm 0.006 & 0.200\pm 0.012 &  -0.335\pm 0.015 &  -0.862\pm 0.008 \\
   -0.031\pm 0.001 &  -0.048\pm 0.001 &  -0.022\pm 0.001 & 0.041\pm 0.001 \\
    0.120\pm 0.0061 & -0.009\pm 0.002 & -0.035\pm 0.002 & -0.126\pm 0.001 \\
    0.970\pm 0.006 & 0.196\pm 0.011 &  -0.341\pm 0.015 &  -0.852 \pm 0.008 
\end{array}\right)$
&  $3.95$  \\ \\ 
$n = 2$ & $\left(\begin{array}{cccc}
    1.000\pm 0.007 &  -0.005\pm 0.013 &  -0.043\pm 0.013 &   0.99\pm 0.02 \\
   -0.017\pm 0.001 &   0.016\pm 0.001 &   0.017\pm 0.001 &  -0.014\pm 0.001 \\
    0.111\pm 0.001 &   0.016\pm 0.001 &  -0.019\pm 0.002 &   0.105\pm 0.002 \\
   -0.994\pm 0.007 &   0.007\pm 0.013 &   0.041\pm 0.013 &  -0.98\pm 0.02
\end{array}\right)$
&  $4870$  \\  \\ 
$n = -2$ & $\left(\begin{array}{cccc}
    1.000\pm 0.006 &  -0.047\pm 0.013 &  0.082\pm 0.012 &  -0.996\pm 0.002\\
    0.003\pm 0.001 &   0.025\pm 0.001 &   0.023\pm 0.001 &  -0.002\pm 0.001\\
   -0.095\pm 0.001 &   0.024\pm 0.002 &  -0.024\pm 0.001 &   0.094\pm 0.001\\
    0.994\pm 0.006 &  -0.045\pm 0.013 &   0.080\pm 0.012 &  -0.991\pm 0.008
\end{array}\right)$
&  $4890$  \\ \\
\end{tabular} 
\end{ruledtabular}
\end{table}
\begin{table}[!ht]
\caption{Mueller matrices measured for each diffraction order (continuation)}
  \label{Mn34}
  \centering
  \begin{ruledtabular}
\begin{tabular}{ccc}
Diffraction order $n$ & Normalized Mueller matrix $\widehat{M}_n/m_{00}^{(n)}$  & $m_{00}^{(n)} \times 10^4$\\
\hline
\\
$n = 3$ & $\left(\begin{array}{cccc}
    1.000\pm 0.007 &   0.009\pm 0.013 &  -0.083\pm 0.014 &   0.89\pm 0.02\\
   -0.054\pm 0.001 &   0.011\pm 0.001 &   0.017\pm 0.001 &  -0.047\pm 0.001\\
    0.096\pm 0.001 &   0.001\pm 0.001 &  -0.018\pm 0.001 &   0.086\pm 0.002\\
   -0.967\pm 0.007 &   0.003\pm 0.013 &   0.092\pm 0.014 &  -0.89\pm 0.02
\end{array}\right)$
&  $3.34$ \\ \\ 
$n = -3$ & $\left(\begin{array}{cccc}
    1.000\pm 0.006 &   0.019\pm 0.013 &   0.188\pm 0.03 &  -0.97\pm 0.03\\
    0.039\pm 0.001 &   0.049\pm 0.001 &   0.029\pm 0.001 &  -0.036\pm 0.001\\
   -0.084\pm 0.001 &   0.031\pm 0.001 &  -0.009\pm 0.001 &   0.079\pm 0.001\\
    0.934\pm 0.006 &   0.004\pm 0.012 &   0.155\pm 0.011 &  -0.914\pm 0.008
\end{array}\right)$
&  $3.75$ \\ \\ 
$n = 4$ & $\left(\begin{array}{cccc}
    1.000\pm 0.007  & -0.87\pm 0.02 &  -0.043\pm 0.013 &  -0.026\pm 0.012\\
   -0.088\pm 0.002  &  0.198\pm 0.004 &  -0.249\pm 0.003 &  -0.272\pm 0.004\\
   -0.219\pm 0.002  &  0.148\pm 0.005 &   0.234\pm 0.005 &  -0.248\pm 0.005\\
   -0.868\pm 0.006  &  0.94\pm 0.02 &   0.061\pm 0.012 &   0.097\pm 0.011
\end{array}\right)$
&  $7.67 $ \\  \\
$n = -4$ & $\left(\begin{array}{cccc}
    1.000\pm 0.006 &   0.230\pm 0.011 &  -0.263\pm 0.015 &   0.373\pm 0.015\\
   -0.002\pm 0.002 &   0.234\pm 0.004 &  -0.533\pm 0.006 &  -0.380\pm 0.005\\
   -0.078\pm 0.002 &   0.543\pm 0.007 &   0.385\pm 0.006 &  -0.174\pm 0.004\\
    0.622\pm 0.005 &   0.412\pm 0.006 &  -0.300\pm 0.011 &   0.625\pm 0.014
\end{array}\right)$
&  $0.650$ \\ \\
\end{tabular} 
\end{ruledtabular}
\end{table}
   
\begin{table}[!ht]
\caption{Mueller matrices measured for each diffraction order (continuation)}
  \label{Mn46}
  \centering
  \begin{ruledtabular}
\begin{tabular}{ccc}
Diffraction order $n$ & Normalized Mueller matrix $\widehat{M}_n/m_{00}^{(n)}$  & $m_{00}^{(n)} \times 10^4$ \\
\hline
\\
$n = 5$ & $\left(\begin{array}{cccc}
    1.000\pm 0.006 &  -0.213\pm 0.015 &   0.244\pm 0.011 &   0.398\pm 0.015\\
   -0.161\pm 0.001 &   0.136\pm 0.004 &  -0.043\pm 0.002 &  -0.170\pm 0.004\\
    0.106\pm 0.001 &   0.043\pm 0.001 &   0.196\pm 0.002 &   0.034\pm 0.002\\
   -0.719\pm 0.005 &   0.228\pm 0.011 &  -0.207\pm 0.008 &  -0.381\pm 0.012
\end{array}\right)$
&  $0.051$ \\ \\
$n = -5$ & $\left(\begin{array}{cccc}
    1.000\pm 0.006 &   0.026\pm 0.013  & -0.110\pm 0.013 &  -0.422\pm 0.009\\
   -0.014\pm 0.001 &  -0.047\pm 0.001  & -0.104\pm 0.001 &  -0.018\pm 0.001\\
   -0.074\pm 0.001 &   0.079\pm 0.002  &  0.116\pm 0.002 &  -0.008\pm 0.001\\
    0.617\pm 0.004 &  -0.001\pm 0.008 &  -0.175\pm 0.009 &  -0.379\pm 0.005
\end{array}\right)$
&  $0.037$ \\ \\
$n = 6$ & $\left(\begin{array}{cccc}
    1.000\pm 0.007 &   0.763\pm 0.009 &  -0.224\pm 0.015 &   0.48\pm 0.02\\
    0.366\pm 0.003 &   0.217\pm 0.004 &  -0.112\pm 0.006 &   0.343\pm 0.008\\
    0.233\pm 0.002 &   0.219\pm 0.002 &   0.118\pm 0.002 &   0.164\pm 0.005\\
   -0.833\pm 0.006 &  -0.709\pm 0.007 &   0.224\pm 0.013 &  -0.392\pm 0.014
\end{array}\right)$
&  $0.371$ \\ \\
$n = -6$ & $\left(\begin{array}{cccc}
    1.000\pm 0.007  & -0.72\pm 0.02 &  -0.004\pm 0.013 &  -0.528\pm 0.009\\
   -0.522\pm 0.003  &  0.325\pm 0.010 &   0.012\pm 0.007 &   0.498\pm 0.005\\
    0.031\pm 0.001  & -0.021\pm 0.003 &   0.285\pm 0.003 &  -0.024\pm 0.001\\
    0.761\pm 0.005  & -0.72\pm 0.02 &  -0.013\pm 0.010 &  -0.317\pm 0.008
\end{array}\right)$
&  $0.476$ \\ \\
\end{tabular}
\end{ruledtabular}
\end{table}

\bibliography{PG}

\end{document}